\documentclass[prb,aps,twocolumn,amsmath,superscriptaddress]{revtex4}

\usepackage{graphicx}
\usepackage{epsfig}
\usepackage{gensymb}
\begin{document}
\title{Four-wave mixing in perovskite photovoltaic materials reveals long dephasing times and weaker many-body interactions than GaAs}

\author{Samuel A. March}
\affiliation{Department of Physics and Atmospheric Science,
Dalhousie University, Halifax, Nova Scotia B3H 4R2 Canada}

\author{Drew B. Riley}
\affiliation{Department of Physics and Atmospheric Science,
Dalhousie University, Halifax, Nova Scotia B3H 4R2 Canada}

\author{Charlotte Clegg}
\affiliation{Department of Physics and Atmospheric Science,
Dalhousie University, Halifax, Nova Scotia B3H 4R2 Canada}

\author{Daniel Webber}

\affiliation{Department of Physics and Atmospheric Science,
Dalhousie University, Halifax, Nova Scotia B3H 4R2 Canada}

\author{Xinyu Liu}

\affiliation{Department of Physics, University of Notre Dame,
Notre Dame, IN 46556}

\author{Margaret Dobrowolska}

\affiliation{Department of Physics, University of Notre Dame,
Notre Dame, IN 46556}

\author{Jacek K. Furdyna}

\affiliation{Department of Physics, University of Notre Dame,
Notre Dame, IN 46556}

\author{Ian G. Hill}
\affiliation{Department of Physics and Atmospheric Science,
Dalhousie University, Halifax, Nova Scotia B3H 4R2 Canada}

\author{Kimberley C. Hall}

\affiliation{Department of Physics and Atmospheric Science,
Dalhousie University, Halifax, Nova Scotia B3H 4R2 Canada}

\begin{abstract}
\bf{Perovksite semiconductors have shown promise for low-cost solar cells, lasers and photodetectors, yet their fundamental photophysical properties are not well understood. Recent observations of a low ($\sim$few meV) exciton binding energy and evidence of hot phonon effects in the room temperature phase suggest that perovskites are much closer to inorganic semiconductors than the absorber layers in traditional organic photovoltaics, signaling the need for experiments that shed light on the placement of perovskite materials within the spectrum of semiconductors used in optoelectronics and photovoltaics.  Here we use four-wave mixing (FWM) to contrast the coherent optical response of CH$_3$NH$_3$PbI$_3$ thin films and crystalline GaAs.  At carrier densities relevant for solar cell operation, our results show that carriers interact surprisingly weakly via the Coulomb interaction in perovskite, much weaker than in inorganic semiconductors.  These weak many-body effects lead to a dephasing time in CH$_3$NH$_3$PbI$_3$ $\sim$3 times longer than in GaAs.  Our results also show that the strong enhancement of the exciton FWM signal tied to excitation-induced dephasing in GaAs and other III-V semiconductors does not occur in perovskite due to weak exciton-carrier scattering interactions.}
\end{abstract}

\pacs{}

\maketitle

Since the first integration of organo-lead trihalide perovskites into photovoltaic devices,\cite{Kojima:2009} the efficiencies of solar cells using this material as the primary absorber layer have increased at an unprecedented rate, having reached over 20\% in just a few years.\cite{NREL:web}   Progress in understanding the fundamental physical properties of these materials has been much slower to develop as the organic-inorganic pervskites are much more complex than both the typical organic semiconductors used in solution-processed photovolatics  and traditional inorganic semiconductor solar cell materials.  The relative importance of excitons and free carriers to the optical response and carrier transport in perovskite systems has been the subject of considerable controversy,\cite{Tanaka:2003,Even:exciton2014,Hirasawa:1994,Huang:2013,Menedez:2014,DInnocenzo:2014,Lin:2014,Savenije:2014,Sun:2015,Miyata:2015,Yamada:2015,Cooke:2015} although a consensus is now emerging that excitonic effects are weaker than had previously been thought,\cite{Even:exciton2014,Miyata:2015,Yamada:2015,Cooke:2015} with optical phonons and the rotational motion of the CH$_{3}$NH$_{3}^{+}$ cations being identified as an essential contributor to dielectric screening and an associated reduction in the exciton binding energy ($E_b$).\cite{Even:exciton2014} Together with recent experiments showing evidence of phonon bottleneck effects\cite{Price:2015} and the successful interpretation of numerous dynamic optical experiments considering only free carriers,\cite{Stranks:2013,Deschler:2014,Trinh:2015,Saba:2014,Wehrenfennig:2013,Manser:2014,Flender:2015,Price:2015} this suggests that the organo-lead trihalide perovskites are more similar to direct band gap III-V semiconductors like GaAs (characterized by $E_b\sim$ 4~meV) than solution-processed organics in which excitonic effects govern both optical excitation and transport.\cite{Hoppe:2004}  Nevertheless, many open questions remain regarding the fundamental photophysical properties of these promising photovoltaic materials.  

Here we contrast the \emph{coherent} carrier dynamics in solution-processed perovskite films with the inorganic semiconductor GaAs using the powerful spectroscopic technique of femtosecond four-wave mixing, providing insight into where perovskite semiconductors fit into the broader landscape of materials used in photovoltaic technology.  In a FWM experiment, two laser pulses $\vec{E_1}(t)$ and $\vec{E_2}(t-\tau_d)$ with wave vectors $\vec{k_1}$ and $\vec{k_2}$ excite a third-order polarization that emits in the direction 2$\vec{k_2}-\vec{k_1}$ (see Fig.~\ref{fig:experiment}, and Supplementary Information).  The magnitude of the emitted signal, often referred to as \emph{self diffraction}, versus the delay time $\tau_d$ probes the decay of quantum coherence within the system of electron-hole pairs excited by the laser pulse, providing a direct measurement of the time scale tied to the strongest scattering process following optical excitation.  Interactions within the system of excited carriers and/or excitons lead to additional contributions to the self-diffraction signal, making this technique especially sensitive to many-body effects.\cite{ShahBook}  As a result of this high sensitivity, fundamental interactions such as electron-electron scattering or exciton-free carrier scattering, which together with electron-phonon coupling govern the relaxation, drift and diffusion of carriers following optical excitation in an operating device, may be studied at low carrier densities reflective of solar cell operating conditions.   

While evidence of many-body effects such as Auger recombination and band gap renormalization have been observed in perovskites at large carrier densities $\gtrsim$1$\times$10$^{18}$~cm$^{-3}$,\cite{Trinh:2015,Price:2015,Wu:2015} our experiments demonstrate that at densities relevant for solar cell device operation Coulomb-mediated scattering plays a negligible role in carrier dephasing, indicating that charge carriers in CH$_3$NH$_3$PbI$_3$ interact surprisingly weakly.  This finding represents an unexpected divergence from the inorganic semiconductors such as GaAs, in which such many-body effects contribute at densities as low as 10$^{13}$ cm$^{-3}$, and are stronger than all competing interactions for temperatures up to 300~K and carrier densities down to 10$^{15}$ cm$^{-3}$.\cite{Schultheis:1986,Cundiff:1996,ElSayed:1997,Hall:2002,Rappen:1994,Shacklette:2002}  In addition to causing a three-fold larger 10~K interband dephasing time relative to GaAs, our results show that the weak many body effects in perovskite have the consequence that the dramatic enhancement of the exciton FWM signal in GaAs tied to excitation-induced dephasing does not occur.  These weak many-body effects highlight a fundamental difference in the nature of charge carrier scattering in perovksite relative to inorganic semiconductors, with crucial implications for both understanding charge dynamics following optical excitation and for optimizing solar cell performance.

\begin{flushleft}
{\bf Weak exciton-carrier scattering in CH$_3$NH$_3$PbI$_3$} \\
\end{flushleft}
The four-wave mixing response of the CH$_3$NH$_3$PbI$_3$ film is shown in Fig.~\ref{fig:tuning200K}(a), illustrating the variation of the signal characteristics as the center photon energy of the laser pulse is tuned relative to the band gap energy.  For these results, the perovskite film was held at 200~K in a liquid N$_2$ cryostat.  The corresponding results for GaAs at 200~K are shown in Fig.~\ref{fig:tuning200K}(b).  Despite a similar estimated exciton binding energy in the room temperature phase of CH$_3$NH$_3$PbI$_3$ and GaAs,\cite{Even:exciton2014,Miyata:2015,Yamada:2015} as well as a similar Urbach energy,\cite{deWolf:2014} the coherent response from the two materials differs dramatically.  In GaAs, the FWM signal consists of a prominent exciton peak, and a broad peak at higher energy associated with unbound electron-hole pairs excited on optical transitions above the band gap.  In contrast, for the pervoskite sample the FWM signal is smooth and featureless for all laser tuning conditions:  Only the overall magnitude of the response varies with laser tuning (Fig.~\ref{fig:tuning200K}(c)).  The spectral dependence of the FWM signal from the perovskite film at 200~K is similar to the laser excitation spectrum (Supplementary Fig. S4), indicating only a free carrier response. The results of FWM experiments on the perovskite film at 10~K are shown in Fig.~\ref{fig:LTcomp10K}(a).  In addition to the interband response above the band gap, a weak shoulder below the band gap is apparent and attributed to the FWM response of the exciton.\cite{March:exciton}  FWM results on GaAs at 10~K are shown in Fig.~\ref{fig:LTcomp10K}(c).  While in perovskite the exciton and interband signals are comparable in magnitude, in GaAs a giant exciton FWM signal is detected.  The large exciton signal measured in GaAs exceeds the interband response by a factor of approximately 15.   

The large exciton signal observed in the FWM response of GaAs, which contrasts with the linear absorption spectrum at 200~K that shows no discernible exciton (Fig.~\ref{fig:tuning200K}(d)), has been studied extensively and is tied to scattering between the excitons and free carriers mediated by the Coulomb interaction (see Supplementary Information for more details).\cite{Schultheis:1986,Cundiff:1996,ElSayed:1997,Hall:2002,Rappen:1994}  This many-body interaction, typically referred to as \emph{excitation-induced dephasing}, results in an additional FWM signal at the exciton that is proportional to the slope of the density dependent dephasing rate.  This EID signal is much larger than the non-interacting exciton self-diffraction signal ({\textit{i.e.} the response expected from a simple two-level system), allowing the exciton FWM signal to persist in GaAs  to temperatures beyond 250~{\rm K} despite the small exciton binding energy of 4~meV.   This many-body signal is easily distinguished from a non-interacting signal because the width of the exciton peak versus delay is determined by the bandwidth of free carrier transitions excited by the laser pulse rather than the intrinsic dephasing rate of the exciton, a consequence of interference between contributions to the overall exciton diffraction signal from free carriers at different energies.\cite{Cundiff:1996,ElSayed:1997,Hall:2002,Rappen:1994} 

It is important to note that the lack of a strong FWM signal tied to excitons in CH$_3$NH$_3$PbI$_3$ does not imply that excitonic effects (including Sommerfeld enhancement in the vicinity of the band gap) are weak.  Excitonic effects result from the Coulomb interaction between electrons and holes and dictate the ground state excitations of the system.  In contrast, our results provide direct insight into the strength of \emph{many-body} effects, which concern the coupling of electron-hole pairs with each other and lead to a hierarchy of interactions and associated FWM signal contributions.\cite{Stone:2009}  In GaAs, the EID signal at the exciton results from four-particle correlations tied to scattering between bound and/or unbound electron-hole pairs.  The lack of a strong exciton signal in the results of FWM experiments on the perovskite sample indicates that such scattering interactions are weak (or equivalently that the exciton dephasing rate is not dependent on the excited carrier density, corresponding to a negligible EID coefficient), in sharp contrast to GaAs.  

A large exciton tied to the same EID many-body process has been observed in experiments on a range of inorganic semiconductors (GaAs, InGaAs, InP, Ge) and in a variety of situations in such materials (quantum well excitons, magneto-excitons, and excitons tied to the spin-orbit split-off band gap).\cite{Cundiff:1996,ElSayed:1997,Hall:2002,Rappen:1994}  Furthermore, strong exciton-carrier scattering has been observed in these systems at densities as low as 1$\times$10$^{13}$~cm$^{-3}$.\cite{Schultheis:1986}  The weak exciton response in the perovskite thin film suggests that exciton-carrier scattering effects are negligible in this system, representing a fundamental departure from the inorganic semiconductors with respect to many-body interactions, and thus a fundamental difference in the physical processes governing carrier relaxation and transport in these materials.

\begin{flushleft}
{\bf Long interband dephasing time in CH$_3$NH$_3$PbI$_3$: Weak carrier-carrier scattering} \\
\end{flushleft}
The coherence decay time ($\tau$) indicates the time scale associated with the fastest scattering events for either electrons or holes following optical excitation, including potential interactions with defects, phonons or other charge carriers. $\tau$ was extracted from the decay of the FWM signal versus the interpulse delay by fitting the 10~K results for perovskite and GaAs to a standard photon echo response\cite{YT:1979} (see Supplementary Information for more details).  The delay dependence of the FWM signal was analyzed at a detection energy 10~meV above the band gap, yielding the scattering times associated with low-energy unbound electron-hole pairs.   The dephasing time extracted from the results on the perovskite sample in Fig.~\ref{fig:LTcomp10K}(a) is 220~fs, well above the time resolution of the experiments ($\sim$50~fs).  The results for GaAs in Fig.~\ref{fig:LTcomp10K}(c) yield a dephasing time of 60~fs.   The slower coherence decay for electron-hole pairs in CH$_3$NH$_3$PbI$_3$ is surprising given the relatively high density of defects present in the solution-processed film.  The density of defects in similar films has been estimated at 1$\times$10$^{16}$~cm$^{-3}$ to 2$\times$10$^{17}$~cm$^{-3}$,\cite{Stranks:2014,Xing:2014} much larger than in crystalline GaAs for which the defect densities are typically $\lesssim$10$^{14}$ cm$^{-3}$.

The slower carrier dephasing in CH$_3$NH$_3$PbI$_3$ relative to GaAs is due to much weaker Coulomb scattering between free carriers in the perovskite sample, as shown in Fig.~\ref{fig:fitresults}, which shows $\tau$ versus the optically-injected carrier density ($n_{eh}$, evaluated using the measured total absorbed power from both excitation beams, laser repetition rate, sample thickness and measured laser spot size).  The rapid interband dephasing process in GaAs has been well characterized and is governed by strong carrier-carrier scattering, described by a dephasing rate $\frac{1}{\tau}=\frac{1}{\tau_0} +c n_{eh}^{\frac{1}{3}}$, where the $\frac{1}{3}$ exponent describes carrier-carrier scattering in the bulk film (\textit{i.e.} for carriers with 3 degrees of freedom of motion), c is the EID coefficient, and the offset $\frac{1}{\tau_0}$ is due to coupling to phonons.\cite{Hugel:1999} Fig.~\ref{fig:fitresults} shows the measured dephasing time at 10~K from our GaAs film as a function of $n_{eh}$.  For comparison, the results of measurements in bulk GaAs at 300~K from Ref.~[\onlinecite{Hugel:1999}] are also shown in Fig.~\ref{fig:fitresults}.  All of the measured results for GaAs fit well to the 1/3 power law, showing dominant carrier-carrier scattering.  In Ref.~[\onlinecite{Hugel:1999}] the 1/3 power law was found to hold over three orders of magnitude in density, down to 2$\times$10$^{15}$ cm$^{-3}$, and a strong dependence of the dephasing rate on carrier density has been observed in earlier studies in GaAs for $n_{eh}$ as low as 1$\times$10$^{13}$~cm$^{-3}$.\cite{Schultheis:1986}

The dephasing time for the CH$_3$NH$_3$PbI$_3$ sample is independent of $n_{eh}$ over the accessible measurement range down to 4$\times$10$^{15}$ cm$^{-3}$. (The range of accessible carrier density is limited by laser power on the high side and signal to noise and on the low side.)    The 10~K measured value for $\tau$ in GaAs of 60~fs corresponds to a more than 3-fold shorter value in crystalline GaAs than in the solution-processed CH$_3$NH$_3$PbI$_3$ film.  These results indicate that, at the densities relevant for solar cell device operation ($\lesssim$10$^{16}$ cm$^{-3}$), Coulomb-medicated carrier carrier scattering is negligible within the perovskite system, in contrast to GaAs in which such effects strongly dominate.

\begin{flushleft}
{\bf Role of Defects: Comparison to Low-temperature-grown GaAs} \\
\end{flushleft}
A possible explanation for the weak many-body effects we observe in CH$_3$NH$_3$PbI$_3$ is strong carrier localization tied to the high density of defects in the solution-processed film.\cite{Wu:2015}   Recent calculations using density functional theory suggest that the density of deep defects is low in this system, accounting for the long observed diffusion lengths tied to weak non-radiative recombination,\cite{Stranks:2013} but that there is still a high density of shallow point defects tied to methylammonium interstitials and lead vacancies.\cite{Yin:2014}  In addition to these shallow point defects, a shallow spatially-varying potential is present associated with long-range correlated orientations of methylammonium molecules, which results in static disorder at low temperatures and dynamic disorder at room temperature.\cite{Ma:2015}   The shallow local potential fluctuations caused by the above defects and correlated MA orientations can lead to charge localization, which may reduce the strength of Coulomb coupling between charge carriers in CH$_3$NH$_3$PbI$_3$.  

In order to gain insight into the potential role of defects in reducing many-body effects, here we contrast the FWM responses of the GaAs and perovskite samples with corresponding measurements on a companion GaAs sample grown at a lower substrate temperature (250$^{\circ}$C), for which defects are intentionally incorporated.  The resulting material is commonly referred to as low-temperature-grown GaAs (LT-GaAs), and has been studied extensively over the past two decades due to its applicability to THz sources and detectors.\cite{Krotkus:2010}  Growth at 250$^{\circ}$C leads to excess As, including a large density ($\sim$1$\times$10$^{19}$ cm$^{-3}$) of As$_{Ga}$ antisite defects and As clusters that cause both mid-gap trap states and local potential fluctuations.\cite{Krotkus:2010} 

The results of FWM experiments on the LT-GaAs sample are shown in Fig.~\ref{fig:LTcomp10K}(b).  The FWM spectrum consists of a distinct exciton peak and a broadband response associated with the transitions above the band gap, similar to the results in GaAs (Fig.~\ref{fig:LTcomp10K}(c)) albeit with a weaker fractional response from the exciton in comparison to the interband transitions and a broader exciton response reflecting the faster dephasing rate in LT-GaAs.  A shoulder is also observed for photon energies below the exciton in LT-GaAs, which has been attributed to the Urbach band tail tied to As disorder.\cite{Webber:2014}  The key observation for the purposes of this work is that the exciton feature remains in the FWM response of LT-GaAs despite the large amount of disorder.  Recent prepulse four-wave mixing experiments have confirmed that the observed exciton peak in LT-GaAs is caused by exciton-carrier scattering.\cite{Webber:2015}  The dephasing process of free carriers is also dominated by carrier-carrier scattering in LT-GaAs, as shown in Fig.~\ref{fig:fitresults}. 

The dominant influence of many-body effects on the carrier kinetics in LT-GaAs mimics the situation in GaAs, suggesting that the high density of defects in the LT-GaAs film does not strongly reduce interactions between charge carriers.   The origin of the weak Coulomb interactions among carriers in the CH$_3$NH$_3$PbI$_3$ system, which possesses a lower defect density than LT-GaAs by several orders of magnitude, is therefore unclear.  Nevertheless, a role played by disorder cannot be ruled out as carrier localization effects will be a sensitive function of the depth and correlation length of the defect-induced local potential fluctuations.\cite{John:1986}  Both the nature of defects and their influence on charge localization are currently under intense investigation in the methylammonium trihalide perovskite family of materials.\cite{Yin:2014,Ma:2015}  

\begin{flushleft}
{\bf Conclusions and Outlook} \\ 
\end{flushleft}
The emergence of CH$_3$NH$_3$PbI$_3$ for high efficiency solar cell applications has introduced a complex yet urgent materials science challenge in the need to unravel the photophysical properties of these materials, for which the combination of both organic and inorganic constituents has led to unexpected high performance as well as subtle complexities.  This fundamental understanding must be developed in the context of competing technologies based on organic and inorganic solar cell materials.  The scattering processes that occur during carrier relaxation, drift and diffusion are especially crucial to device function.  While incoherent techniques such as pump probe spectroscopy have provided valuable information about carrier cooling and recombination, assessing the relative strength of the operative scattering processes is essential, especially at carrier densities relevant for an operating solar cell device.  Direct access to such scattering process necessitates a \emph{coherent} optical probe that measures the time scale for decay of quantum coherence within the electron-hole pair system, reflecting the transition between the quantum kinetic and thermal regimes following optical excitation of the semiconductor.  

By contrasting the coherent nonlinear optical response of a CH$_3$NH$_3$PbI$_3$ thin film with crystalline GaAs, our experiments illuminate crucial differences between perovskites and inorganic semiconductors used in solar cells and optoelectronics. At densities relevant for solar cell device operation, there is no evidence of many-body effects in perovskite, despite the fact that such effects strongly dominate the carrier kinetics in inorganic group III-V and group IV semiconductors.  In addition, no excitonic feature is observed in the high temperature phase of perovskite and only a weak exciton response is observed in the low temperature phase down to 10~K, in contrast to GaAs which exhibits a prominent excitonic feature at all temperatures up to 250 K. We show that the weaker exciton FWM signal for CH$_3$NH$_3$PbI$_3$ is due to negligible exciton-carrier interactions.  Our experiments highlight the need for further studies to characterize disorder and it's potential influence on carrier localization within the organic-inorganic perovskite solar cell materials.

\begin{flushleft}
{\bf Methods} \\ 
Methods and any associated references are available in the online version of the paper.
\end{flushleft}

\begin{flushleft}
{\bf Acknowledgements} \\ 
This research was supported by the Natural Sciences and Engineering Research Council
of Canada and the Canada Research Chairs Program.  The work at Notre Dame was supported by NSF Grant DMR100432.
\end{flushleft}

\clearpage
Figures.

\begin{figure}[htb]
	\begin{center}
    \includegraphics[width=17 cm]{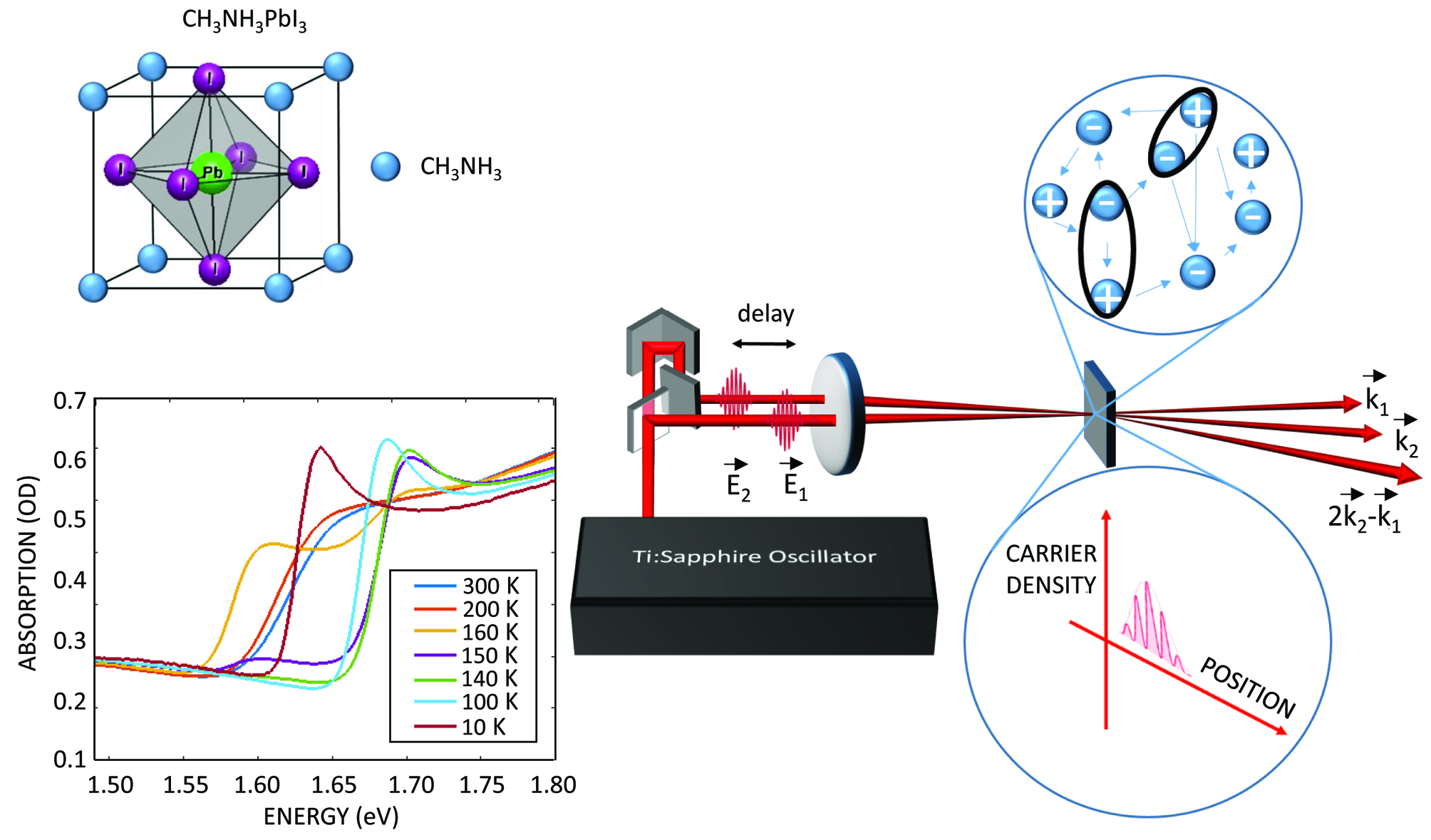}
  \end{center}
    \caption{(Right) {\bf Transient four-wave mixing spectroscopy}:  Interference of $\vec{E_2}$ with the coherent polarization density excited by $\vec{E_1}$ results in a spatially-modulated carrier density in the semiconductor sample. Self-diffraction of $\vec{E_2}$ (non-interacting two-level system) or the polarization density excited by $\vec{E_2}$ (via many-body interactions within the system of electron-hole pairs) constitutes the four-wave mixing signal emitted along 2$\vec{k_2}$-$\vec{k_1}$.  Top left:  Methyl-ammonium lead-iodide perovskite crystal structure. CH$_3$NH$_3$ (blue), Pb (green), I (purple). Bottom left: Absorption spectrum versus temperature, indicating the phase transition around 160 K.\cite{DInnocenzo:2014}}
    \label{fig:experiment}
\end{figure}

\clearpage
\begin{figure}[htb]
	\begin{center}
    \includegraphics[width=17 cm]{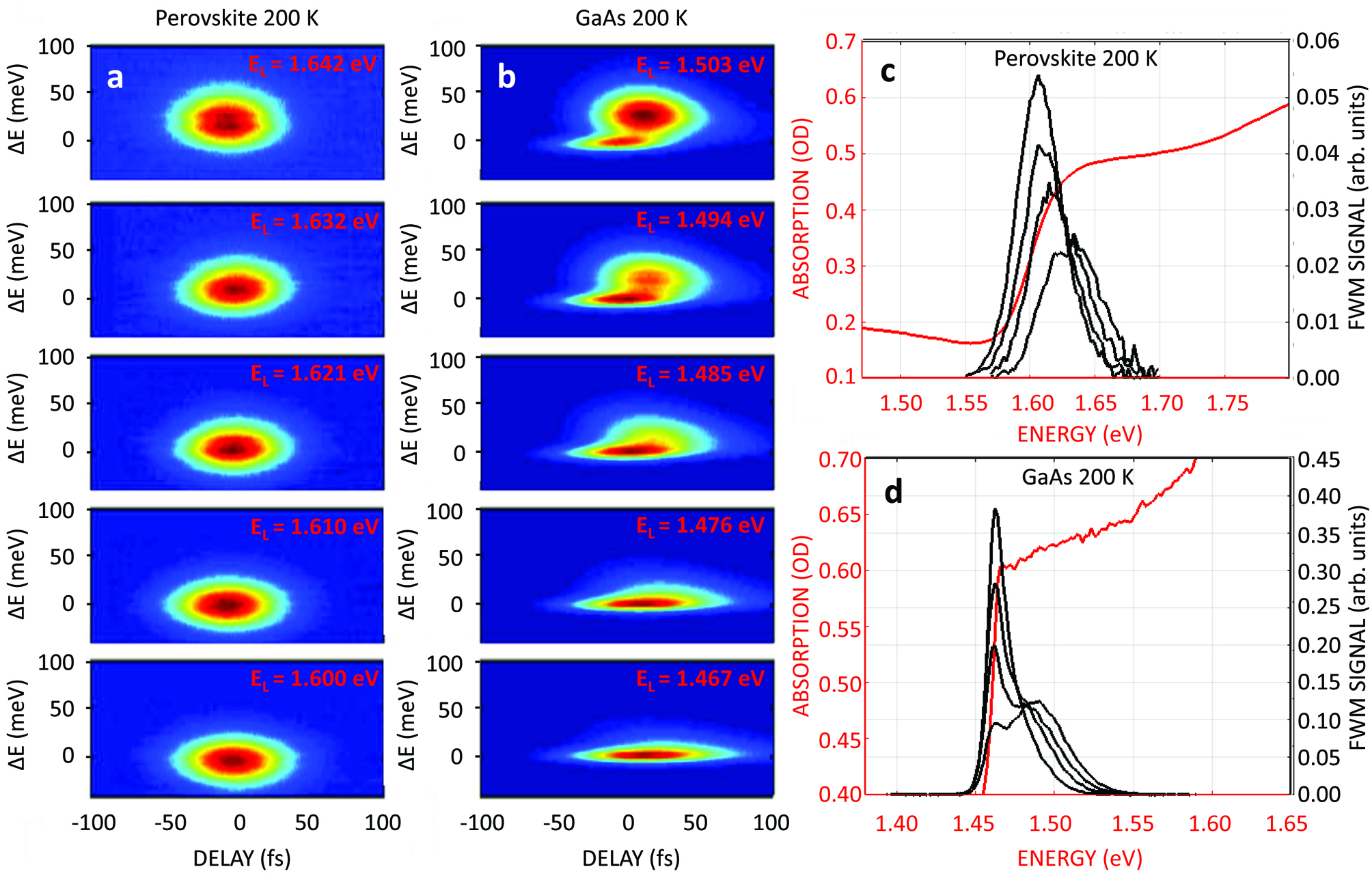}
  \end{center}
    \caption{{\bf Comparison of four-wave mixing response at 200 K from the CH$_3$NH$_3$PbI$_3$ thin film and single crystal GaAs}.  The amplitude of the four-wave mixing signal normalized to the peak value (colour scale) is shown as a function of the time delay between the two excitation laser pulses and the detection energy (E$_{D}$) for different values of the detuning of the laser energy (E$_{L}$) from the band gap energy (E$_g$) ($\Delta$E$\equiv$E$_{D}$-E$_g$). {\bf a} Results on CH$_3$NH$_3$PbI$_3$, with E$_g$ = 1.606~eV, and laser tuning from 1.600 eV (bottom panel) to 1.642 eV (top panel).  {\bf b} Same results for GaAs, with E$_g$ = 1.465 eV and laser tuning from 1.465 eV (bottom panel) to 1.503 eV (top panel).  GaAs shows separate responses from the exciton and free carrier transitions, where the strong exciton peak is caused by Coulomb coupling of the exciton with the free carriers via excitation-induced dephasing. In contrast, the CH$_3$NH$_3$PbI$_3$ shows only a signal from free carrier transitions. {\bf c}  Linear absorption (red) is shown with spectral slices of the contour plots in {\bf a} at zero delay for CH$_3$NH$_3$PbI$_3$.   {\bf d} Same data for GaAs.  For GaAs, the laser tuning determines the relative size of the exciton and free carrier response, whereas in perovksite only the magnitude of the free carrier response varies with tuning.}
    \label{fig:tuning200K}
\end{figure}

\clearpage
\begin{figure}[htb]
	\begin{center}
		\includegraphics[width=17 cm]{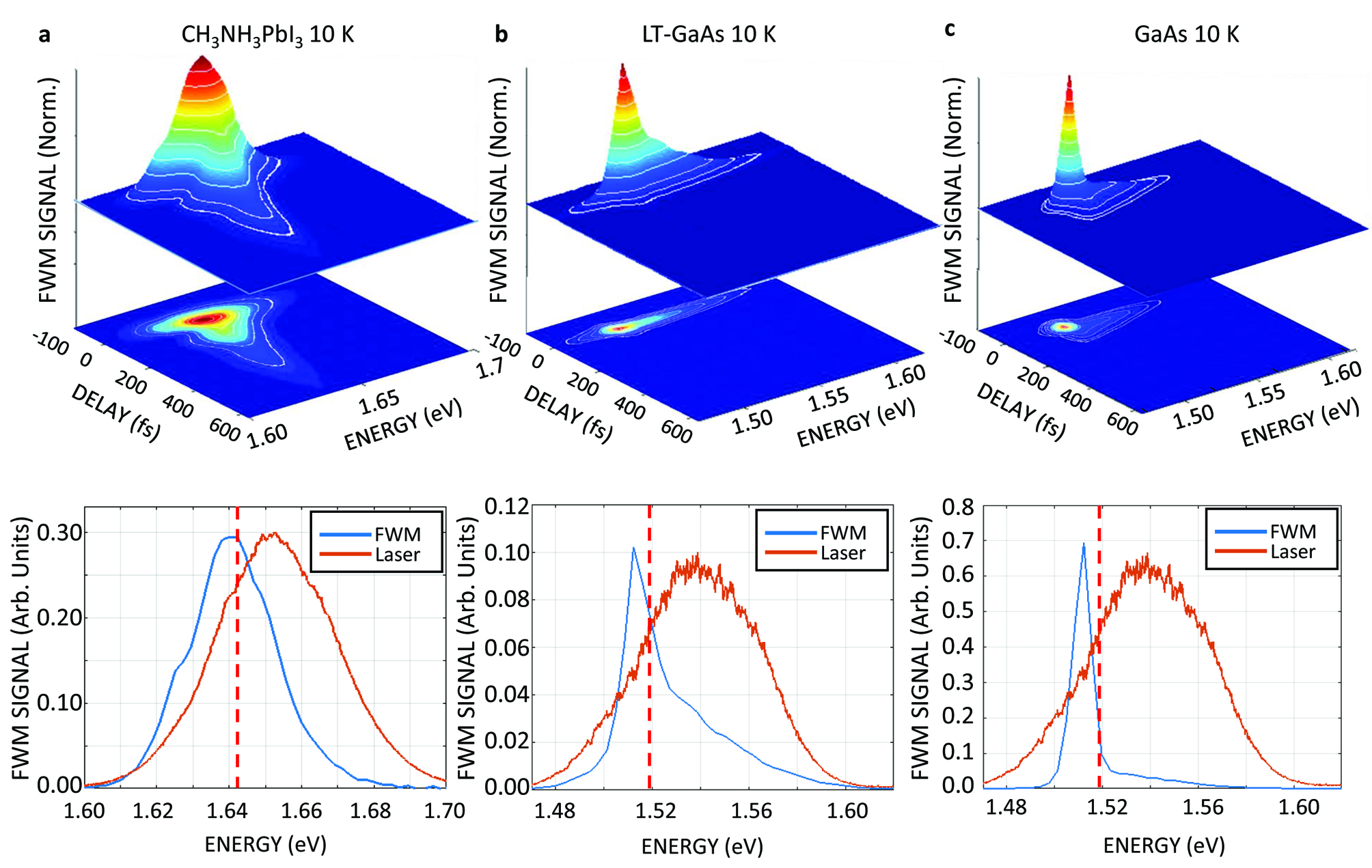}
	\end{center}
    \caption{{\bf Four-wave mixing results for CH$_3$NH$_3$PbI$_3$, low-temperature-grown GaAs, and GaAs at 10~K}.  Upper panels: FWM signal (contour scale) versus pulse delay and photon energy.  Lower panels: spectral cuts at zero delay together with laser spectrum.  The band gap for each sample is indicated by the dashed line.  LT-GaAs and GaAs both show an exciton response caused by exciton-carrier scattering, indicating that defect-induced localization associated with the high density of defects in the LT-GaAs film ($\sim$10$^{19}$~cm$^{-3}$) does not suppress many-body effects.  For the perovskite sample, only a weak exciton signal relative to the interband response is observed and the coherent emission persists to longer time scales than GaAs or LT-GaAs, indicating weak many-body interactions tied to exciton-carrier and carrier-carrier scattering.}
    \label{fig:LTcomp10K}
\end{figure}
\clearpage

\begin{figure}[htb]
	\begin{center}
		\includegraphics[width=8.5 cm]{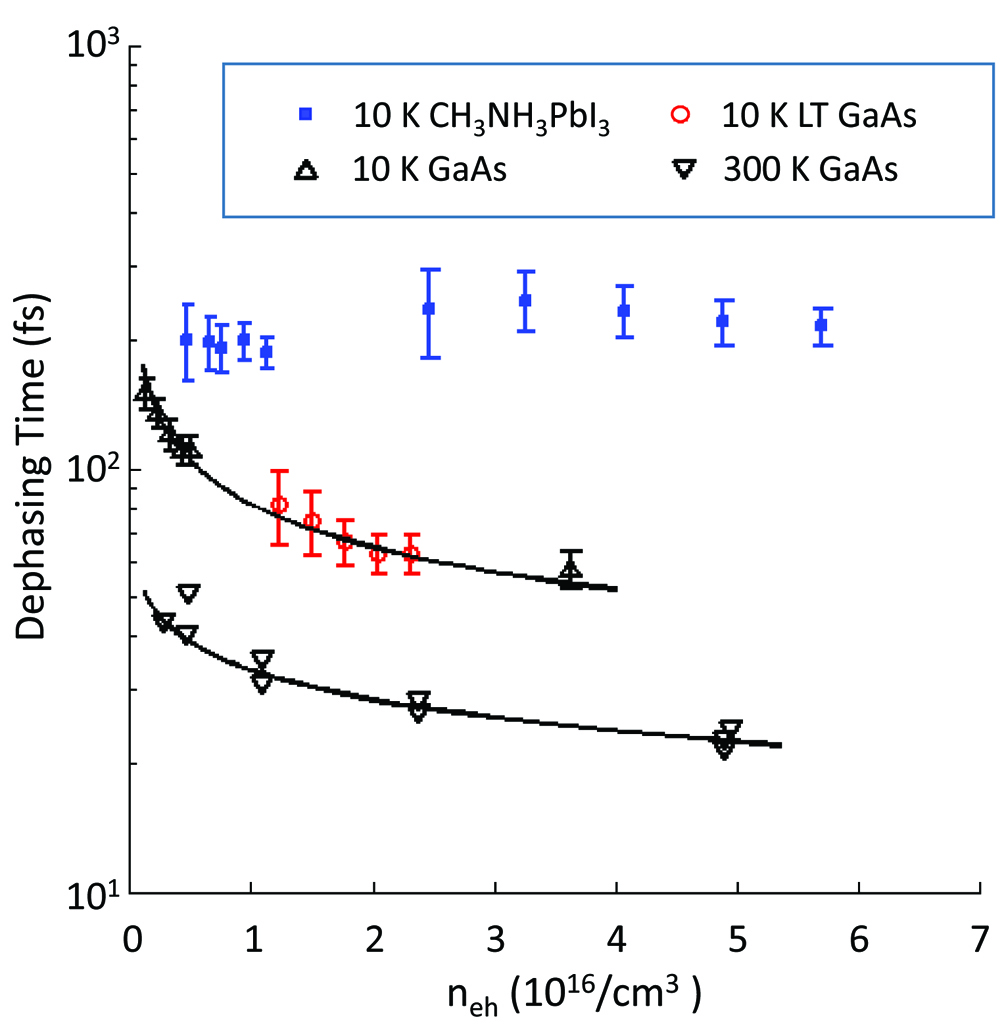}
	\end{center}
    \caption{{\bf Electron-hole pair dephasing times versus the excited carrier density for CH$_3$NH$_3$PbI$_3$, low-temperature-grown GaAs, and GaAs}.   The 10~K results for GaAs and LT-GaAs from Ref.~[\onlinecite{Yildirim:2012}] were taken on the same samples as studied in this work.  The 300~K GaAs experimental data and fit were extracted from Ref.~[\onlinecite{Hugel:1999}].  The dephasing time for perovskite is $>$3 times larger than GaAs under the same conditions and is independent of carrier density, indicating that carrier-carrier scattering does not contribute to dephasing. This contrasts with GaAs, for which carrier-carrier scattering dominates over all other interactions at densities down to 1$\times$10$^{15}$~cm$^{-3}$ and temperatures up to 300~K.\cite{Schultheis:1986,Hugel:1999}}
    \label{fig:fitresults}
\end{figure}
\clearpage




\begin{references}



\bibitem{Kojima:2009} Kojima, A., Teshima, K., Shirai, Y. and Miyasaka, T. Organometal Halide Perovskites as Visible-Light Sensitizers for Photovoltaic Cells. \textit{J. Am. Chem. Soc.} {\bf 131}, 6050-6051 (2009).








\bibitem{NREL:web} http://www.nrel.gov/ncpv/images/efficiency\_chart.jpg.














\bibitem{Even:exciton2014} Even, J., Pedesseau, L. and Katan, C. Analysis of Multivalley and Multibandgap Absorption and Enhancement of Free Carriers Related to Exciton Screening in Hybrid Perovskites. \textit{J. Phys. Chem. C} {\bf 118}, 11566-11572 (2014).

\bibitem{Tanaka:2003} Tanaka, K., Takahashi, T., Ban, T., Kondo, T., Uchida, K., Miura, N. Comparative study on the excitons in lead-halide-based perovskite-type crystals CH$_3$NH$_3$PbBr$_3$ CH$_3$NH$_3$PbI$_3$. \textit{Solid State Commun.}  {\bf 127}, 619-623 (2003). 

\bibitem{Hirasawa:1994}   Hirasawa, M., Ishihara, T., Goto, T., Uchida, K., Miura, N. Magnetoabsorption of the lowest exciton in perovskite-type compound (CH$_3$NH$_3$)PbI$_3$.  \textit{Physica B: Condensed Matter} {\bf 201}, 427-430 (1994).

\bibitem{Huang:2013} Huang, L.-Y., Lambrecht, W. R. L.  Electronic band structure, phonons, and exciton binding energies of halide perovskites CsSnCl$_3$, CsSnBr$_3$, and CsSnI$_3$. \textit{Phys. Rev. B} {\bf 88}, 165203 (2013).

\bibitem{Menedez:2014} Menéndez-Proupin, E., Palacios, P., Wahnon, P., Conesa, J. C.  Self-consistent relativistic band structure of the CH3NH3PbI3 perovskite \textit{Phys. Rev. B} {\bf 90}, 045207 (2014).

\bibitem{DInnocenzo:2014}   D\'{} Innocenzo, V. ,	Grancini, G.,	Alcocer, M. J. P.,	Kandada, A. R. S.,	Stranks, S. D.,	Lee, M. M.,	Lanzani, G.,	Snaith, H. J., Petrozza, A. Excitons versus free charges in organo-lead tri-halide perovskites. Excitons versus free charges in organo-lead tri-halide perovskites. \textit{Nat. Commun.} {\bf 5}, 3586 (2014).

\bibitem{Lin:2014}   Lin, Q.,	Armin, A.,	Nagiri, R. C. R.,	Burn, P. L., Meredith, P.  Electro-optics of perovskite solar cells. \textit{Nat. Photon.} {\bf 9}, 106-112 (2015). 

\bibitem{Savenije:2014}  Savenije, T. J., Ponseca, Jr. C. S., Kunneman, L., Abdellah, M.,  Zheng, K., Tian, Y.,  Zhu, Q., Canton, S. E., Scheblykin, I. G. , Pullerits, T.,  Yartsev, A., and Sundstr\"{o}m, V.,  Thermally Activated Exciton Dissociation and Recombination Control the Carrier Dynamics in Organometal Halide Perovskite \textit{J. Phys. Chem. Lett.} {\bf 5}, 2189-2194 (2014).

\bibitem{Sun:2015} Sun, S., Salim, T., Mathews, N.,  Duchamp, M.,   Boothroyd, C.,  Xing, G., Sum, T. C., Lam, Y. M.   The origin of high efficiency in low-temperature solution-processable bilayer organometal halide hybrid solar cells. \textit{Energy Environ. Sci.} {\bf 7}, 399-407 (2014).

\bibitem{Miyata:2015}  Miyata, A.,	Mitioglu, A.,	Plochocka, P.,	Portugall, O.,	Wang, J. T.-W.,	Stranks, S. D.,	Snaith, H. J., Nicholas R. J. Direct measurement of the exciton binding energy and effective masses for charge carriers in organic–inorganic tri-halide perovskites. \textit{Nature Physics} {\bf 11}, 582-587 (2015).

\bibitem{Yamada:2015} Yamada, Y., Nakamura, T., Endo, M., Wakamiya, A., Kanemitsu, Y. Photoelectronic Responses in Solution-Processed Perovskite CH$_3$NH$_3$PbI$_3$ Solar Cells Studied by Photoluminescence and Photoabsorption Spectroscopy. \textit{IEEE J. Photov.} {\bf 5}, 401-405 (2015).

\bibitem{Cooke:2015} Valverde-Ch\'{a}vez, D. A.,  Ponseca, Jr. C. S.,   Stoumpos, C. C.,  Yartsev, A., Kanatzidis, M. G.,  Sundstr\"{o}m, V., Cooke D. G. Intrinsic femtosecond charge generation dynamics in single crystal CH$_3$NH$_3$PbI$_3$. \textit{Energy Environ. Sci.} {\bf 8}, 3700-3707 (2015).

\bibitem{Price:2015}  Price, M. B.,	Butkus, J.,	Jellicoe, T. C.,	Sadhanala, A.,	Briane, A.,	Halpert, J. E.,	Broch, K.,	 Hodgkiss, J. M.,	Friend, R. H., Deschler, F. Hot-carrier cooling and photoinduced refractive index changes in organic–inorganic lead halide perovskites.  \textit{Nat. Commun.} {\bf 6}, 8420 (2015).

\bibitem{Deschler:2014}  Deschler, F., Price, M., Pathak, S., Klintberg, L. E., Jarausch, D.-D., Higler, R., Huttner, S., Leijtens, T., Stranks, S. D., Snaith, H. J., Atature, M., Phillips, R. T. and Friend, R. H. High Photoluminescence Efficiency and Optically Pumped Lasing in Solution-Processed Mixed Halide Perovskite Semiconductors. \textit{J. Phys. Chem. Lett.} {\bf 5}, 1421-1426 (2014). 


\bibitem{Trinh:2015} Trinh, M. T., Wu, X., Niesner, D., Zhu, X.-Y. Many-body interactions in photo-excited lead iodide perovskite. \textit{J. Mater. Chem. A} {\bf 3}, 9285 (2015).

\bibitem{Saba:2014}  Saba, M.,	Cadelano, M.,	Marongiu, D.,	Chen, F.,	Sarritzu, V.,	Sestu, N.,	Figus, C.,	Aresti, M.,	Piras, R.,	Lehmann, A. G.,	Cannas, C.,	Musinu, A.,	Quochi, F.,	Mura, A., Bongiovanni, G., Correlated electron–hole plasma in organometal perovskites.  \textit{Nat. Commun.} {\bf 5}, 5049 (2014).

\bibitem{Wehrenfennig:2013} Wehrenfennig, C., Eperon, G. E., Johnston, M. B., Snaith, H. J., Herz, L. M.  High Charge Carrier Mobilities and Lifetimes in Organolead Trihalide Perovskites. \textit{Adv. Mater.} {\bf 26}, 1584-1589 (2014).

\bibitem{Manser:2014}  Manser, J. S., Kamat, P. V. Band filling with free charge carriers in organometal halide perovskites.   \textit{Nature Photonics} {\bf 8}, 737–743 (2014).

\bibitem{Flender:2015} Flender, O., Klein, J. R., Lenzer, T., Oum, K. Ultrafast photoinduced dynamics of the organolead trihalide perovskite CH$_3$NH$_3$PbI$_3$ on mesoporous TiO2 scaffolds in the 320–920 nm range. \textit{Phys. Chem. Chem. Phys.} {\bf 17}, 19238-19246 (2015).



\bibitem{Stranks:2013} Stranks, S. D., Eperon, G. E., Grancini, G., Menelaou, C., Alcocer, M. J. P., Leijtens, T.,
Herz, L. M., Petrozza, A., Snaith, H. J.  Electron-Hole Diffusion Lengths Exceeding 1 Micrometer in an Organometal Trihalide Perovskite Absorber. \textit{Science}  {\bf 342}, 341-344 (2013).

\bibitem{Hoppe:2004}  
Hoppe, H., Sariciftci, N. S.  Organic solar cells: An overview. \textit{J. Mater. Res.}, {\bf 19}, 1924-1945 (2004).



\bibitem{ShahBook} Shah, J. \textit{Ultrafast Spectroscopy of Semiconductors and Semiconductor Nanostructures}  (Springer-Verlag, 1996).

\bibitem{Wu:2015} Wu, X., Trinh, M. T., and Zhu, X.-Y. Excitonic Many-Body Interactions in Two-Dimensional Lead Iodide Perovskite Quantum Wells. \textit{J. Phys. Chem. C} {\bf 119}, 14714–14721 (2015).


\bibitem{March:exciton} March, S. A., Clegg, C., Riley, D. B., Webber, D., Hill, I. G. and Hall, K. C. Simultaneous observation of free and defect-bound excitons in CH3NH3PbI3 using four-wave mixing spectroscopy.  arXiv:1608.02019 (2016).




\bibitem{Schultheis:1986} Schultheis, L., Kuhl, J., Honold, A., and Tu, C. W. Ultrafast Phase Relaxation of Excitons via Exciton-Exciton and Exciton-Electron Collisions. \textit{Phys. Rev. Lett.} {\bf 57}, 1635 (1986).

\bibitem{Cundiff:1996} Cundiff, S. T., Koch, M., Knox, W. H., Shah, J., Stolz, W. Optical Coherence in Semiconductors: Strong Emission Mediated by Nondegenerate Interactions. \textit{Phys. Rev. Lett.} {\bf 77}, 1107-1110 (1996).
\bibitem{ElSayed:1997} El Sayed, K., Birkedal, D., Lyssenko, V. G., Hvam, J. M. Continuum contribution to excitonic four-wave mixing due to interaction-induced nonlinearities: A numerical study. \textit{Phys. Rev. B} {\bf 55}, 2456-2465 (1997).
\bibitem{Hall:2002} Hall, K. C., Allan, G. R., van Driel, H. M., Krivosheeva, T., P\"{o}tz, W.  Coherent response of spin-orbit split-off excitons in InP: Isolation of many-body effects through interference. \textit{Phys. Rev. B.} {\bf 65}, 201201(R) (2002).
\bibitem{Rappen:1994} Rappen, T., Peter, U.-G., Wegener, M., Sch\"{a}fer, W. Polarization dependence of dephasing processes: A probe for many-body effects. \textit{Phys. Rev. B} {\bf 49}, 10774(R) (1994).
\bibitem{Stone:2009} Stone, K. W., Turner, D. B., Gundogdu, K., Cundiff, S. T. and Nelson, K. A. Exciton-exciton correlations revealed by two-quantum, two-dimensional Fourier transform optical spectroscopy. \textit{Acc. Chem. Res.} {\bf 42}, 1452 (2009).

\bibitem{Shacklette:2002} Shacklette, J. M., Cundiff, S. T. Role of excitation-induced shift in the coherent optical response of semiconductors. \textit{Phys. Rev. B} {\bf 66}, 045309 (2002).




\bibitem{deWolf:2014} De Wolf, S., Holovsky, J., Moon, S.-J., L\"{o}per, P., Niesen, B., Ledinsky, M., Haug, F.-J., Yum, J.-H., and Ballif, C. Organometallic Halide Perovskites: Sharp Optical Absorption Edge and Its Relation to Photovoltaic Performance. \textit{J. Phys. Chem. Lett.} {\bf 5}, 1035–1039 (2014).

\bibitem{YT:1979} Yajima, T. and Taira, Y. Spatial Optical Parametric Coupling of Picosecond Light Pulses and Transverse Relaxation Effect in Resonant Media. \textit{J. Phys. Soc. Jpn.} {\bf 47}, 1620 (1979).

\bibitem{Xing:2014} Xing, G., Mathews, N., Lim, S. S., Yantara, N., Liu, X., Sabba, D., Gratzel, M., Mhaisalkar, S. and Sum, T. C. Low-temperature solution-processed wavelength-tunable perovskites for lasing. \textit{Nat. Mater.}  {\bf 13}, 476-480 (2014).

\bibitem{Stranks:2014} Stranks, S. D., Burlakov, V. M., Leijtens, T., Ball, J. M., Goriely, A., and Snaith, H. J. Recombination Kinetics in Organic-Inorganic Perovskites: Excitons, Free Charge, and Subgap States.
\textit{Phys. Rev. Applied} {\bf 2}, 034007 (2014).


\bibitem{Hugel:1999}  H\"{u}gel, W. A., Heinrich, M. F., Wegener, M., Vu, Q. T., Banyai, L., and Haug, H. Photon Echoes from Semiconductor Band-to-Band Continuum Transitions in the Regime of Coulomb Quantum Kinetics.
\textit{Phys. Rev. Lett.} {\bf 83}, 3313 (1999).



\bibitem{Yin:2014}  Yin, W.-J., Shi, T, and Yan, Y. Unusual defect physics in CH$_3$NH$_3$PbI$_3$ perovskite solar cell absorber. \textit{Appl. Phys. Lett.} {\bf 104}, 063903 (2014)

\bibitem{Ma:2015} Ma, J.and Wang, L.-W. Nanoscale Charge Localization Induced by Random Orientations of Organic Molecules in Hybrid Perovskite CH$_3$NH$_3$PbI$_3$. \textit{Nano Lett.} {\bf 15} 248–253 (2015).

\bibitem{Krotkus:2010} Krotkus, A. Semiconductors for terahertz photonics applications. \textit{J. Phys. D: Appl. Phys.} {\bf 43}, 273001 (2010).

\bibitem{Webber:2014}  Webber, D., Yildirim, M., Hacquebard, L., March, S., Mathew, R., Gamouras, A., Liu, X., Dobrowolska, M., Furdyna, J. K. and Hall, K. C.  Observation of the exciton and Urbach band tail in low-temperature-grown GaAs using four-wave mixing spectroscopy. \textit{Appl. Phys. Lett.} {\bf 105}, 182109 (2014).

\bibitem{Webber:2015}  Webber, D., Hacquebard, L., Liu, X., Dobrowolska, M., Furdyna, J. K. and Hall, K. C. Role of many-body effects in the coherent dynamics of excitons in low-temperature-grown GaAs. \textit{Appl. Phys. Lett.} {\bf 107}, 142108 (2015).


\bibitem{John:1986} John, S., Soukoulis, C., Cohen, M. H., and Economou, E. N. Theory of Electron Band Tails and the Urbach Optical-Absorption Edge. \textit{Phys. Rev. Lett.} {\bf 57}, 1777 (1986).

\bibitem{Yildirim:2012}  Yildirim, M., March, S., Mathew, R., Gamouras, A., Liu, X., Dobrowolska, M., Furdyna, J. K. and Hall, K. C. Interband dephasing and photon echo response in GaMnAs. \textit{Appl. Phys. Lett.} {\bf 101}, 062403 (2012).

\end{references}
\end{document}